\documentclass{aa}
\usepackage[utf8]{inputenc}
\usepackage[OT2,T1]{fontenc}
\usepackage{textcomp}

\begin{document}

\title{The [CII] and FIR properties of  $z>6$ radio-loud quasars}

\author{Y.~Khusanova\inst{\ref{MPIA}}
\and E.~Ba\~nados\inst{\ref{MPIA}}
\and C.~Mazzucchelli\inst{\ref{ESOChile}}
\and S.~Rojas-Ruiz\inst{\ref{MPIA}}\thanks{Fellow of the International Max Planck Research School for Astronomy and Cosmic Physics at the University of Heidelberg (IMPRS-HD).}
\and E.~Momjian\inst{\ref{NRAO}}
\and F.~Walter\inst{\ref{MPIA}}
\and R.~Decarli\inst{\ref{BolognaObs}}
\and B.~Venemans\inst{\ref{Leiden}}
\and E.~P.~Farina\inst{\ref{Gemini}}
\and R.~Meyer\inst{\ref{MPIA}}
\and F.~Wang\inst{\ref{StewardObs}}
\and J.~Yang\inst{\ref{StewardObs}}
}
\institute{Max-Planck-Institut f\"{u}r Astronomie, K\"{o}nigstuhl 17, D-69117 Heidelberg, Germany\label{MPIA}
\and European Southern Observatory, Alonso de Cordova 3107, Vitacura, Region Metropolitana, Chile\label{ESOChile}
\and National Radio Astronomy Observatory, PO Box O, Socorro, NM 87801, USA\label{NRAO}
\and INAF - Osservatorio di Astrofisica e Scienza dello Spazio di Bologna, via Gobetti 93/3, I-40129, Bologna, Italy\label{BolognaObs}
\and Leiden Observatory, Leiden University, PO Box 9513, 2300 RA Leiden, The Netherlands\label{Leiden}
\and Gemini Observatory, NSF’s NOIRLab, 670 N A’ohoku Place, Hilo, Hawai'i 96720, USA\label{Gemini}
\and Steward Observatory, University of Arizona, 933 N. Cherry Ave., Tucson, AZ 85721, USA\label{StewardObs}
}

\abstract{There are only five radio-loud quasars currently known within 1 Gyr from the Big Bang ($z>6$) and the properties of their host galaxies have not been explored in detail. 
We present a NOrthern Extended Millimeter Array (NOEMA) survey of [CII] (158 $\mu$m) and underlying continuum emission of four $z>6$ radio-loud quasars, revealing their diverse properties. J0309+2717 ($z=6.10$) has a bright [CII] line and underlying continuum, implying a starburst with a star-formation rate SFR=340--1200\,$M_\odot$\,yr$^{-1}$.
J1429+5447 ($z=6.18$) has a  SFR=$520-870$\,$M_{\odot}$yr$^{-1}$ and its [CII] profile is consistent with two Gaussians, which could be  interpreted as a galaxy merger.
J1427+3312 ($z=6.12$) has a moderate SFR = 30--90\,$M_\odot$\,yr$^{-1}$. Notably, this is a broad absorption line quasar and we searched for the presence of high-velocity outflows in the host galaxy. Although the NOEMA data reveal a tentative broad component of the [CII] line as wide as $\sim$1400~km~s$^{-1}$, the sensitivity of our current data is not sufficient to confirm it. 
Finally, P172+18 ($z=6.82$) is undetected in both [CII] and continuum, implying a SFR$<22-40$\,$M_{\odot}$yr$^{-1}$.  The broad range of SFRs is similar to what is observed in radio-quiet quasars at similar redshifts. If radio jets do not significantly contribute to both [CII] and IR luminosities, this suggest no feedback from the jet on the star formation in the host galaxy.}


\titlerunning{The [CII] and FIR properties of  $z>6$ radio-loud quasars}
\authorrunning{Yana Khusanova et al.}

\maketitle

\section{Introduction}

The formation and evolution of supermassive black holes is one of the key unresolved puzzles in astrophysics. 
Black holes (BHs) of masses $\sim10^9 M_{\odot}$ are already in place in the first Gyr after the Big Bang \citep[e.g.,][]{Banados2018, Yang2020, Wang2021}. 
It is still a mystery how these BHs acquire such masses and how their evolution is connected to that of the host galaxies \citep[e.g.,][]{Inayoshi2020,Volonteri2021}. 
The mergers of host galaxies can lead to a coalescence of their respective BHs, while simultaneously playing a role in driving the gas towards the center, thus enabling efficient accretion onto the BH. This can result in the formation of relativistic jets \citep[e.g.][]{Chiaberge2015}. 
The presence of a jet allows to partially convert accretion power to non-radiative form, thus enhancing the accretion rate \citep[e.g.,][]{Jolley2008}. 
Hence, mergers of host galaxies and possible subsequent formation of jets can play a crucial role in the fast growth of BHs in the early Universe \citep{Volonteri2015}.

The presence of jets can affect the evolution of the host galaxies in different ways. 
While AGN-driven extreme gas outflows can lead to quenching of star formation in the host galaxy \citep[e.g.,][]{DiMatteo2005,Villar2014}, these jets can also trigger enhanced star formation in the host galaxy via AGN-induced pressure \citep[e.g.,][]{Silk2013}. 
It is unclear which mechanism plays the most important role at high redshift, since no systematic study of host galaxies of quasars with jets at $z\gtrsim6$ has been undertaken. Relativistic jets produce synchrotron radio emission. Therefore, radio-loudness\footnote{$R_{4400}=S_{5GHz}/S_{4400\text{\normalfont\AA}}$ , where $S_{5GHz}$ and $S_{4400\AA}$ are flux densities at the rest-frame at 5\,GHz and 4400\,\AA, respectively \citep{Kellermann1989}} $R_{4400}>10$ is a sign of the existence of such jets. Such quasars are considered radio-loud and quasars with $R_{4400}<10$ are called radio-quiet.

The host galaxies of radio-quiet quasars at $z\gtrsim6$ have been extensively observed using the Atacama Large Millimeter/submillimeter Array (ALMA) and the NOrthern Extended Millimeter Array (NOEMA) revealing the cold dust emission in the rest-frame far-infrared (FIR) and [CII] line emission \citep[e.g.][]{Wang2013, Willott2015, Banados2015, Willott2017,Decarli2018, Izumi2018, Izumi2019, Venemans2020}. 
These studies showed that the host galaxies of radio-quiet quasars have star formation rates (SFRs) reaching up to 2500 M$_{\odot}$/yr \citep[e.g.,][]{Venemans2020} and one third shows signs of recent mergers \citep{Neeleman2021}. Many studies have sought out to confirm active galactic nuclei (AGN) driven outflows in the host galaxies. However, the results are still inconclusive \citep[][]{Cicone2015,Bischetti2019, Novak2020, Meyer2022}. Moreover, it is unknown whether the host galaxies of radio-loud quasars differ from those of radio-quiet ones, whether they have outflows, how common are mergers among them and how the radio-mode AGN feedback affects star formation at $z>6$.

Radio-loud quasars are relatively rare as they constitute only 8-10\% of quasar population at $z\gtrsim6$ \citep{Banados2015_rlq_fraction, Liu2021, Gloudemans2021}. 
To date there are only five radio-loud quasars known at $z>6$:  J1427+3312 at $z=6.12$ \citep{McGreer2006}, J1429+5447 at $z=6.18$ \citep{Willott2010}, J0309+2717 at $z=6.10$ \citep{Belladitta2020}, J2318--3113 at $z=6.44$ \citep{Ighina2021} and P172+18 at $z=6.83$ \citep{Banados2021}. 
Only the rest-frame FIR properties of one of them, J2318--3113, have been studied before this source was recognized as radio-loud \citep{Decarli2018,Venemans2020,Neeleman2021}.

In this paper, we report NOEMA observations of [CII] emission and the underlying continuum for the remaining 4 radio-loud quasars known at $z>6$. 
In Section~\ref{sect:data}, we describe the observations and data reduction. 
In Section~\ref{sect:results}, we present the results for individual objects and discuss their properties. 
In Section~\ref{sect:discussion}, we compare the [CII] emission and dust properties of all currently known $z\gtrsim6$ radio-loud quasars with the samples of radio-quiet quasars at $z\gtrsim6$ from the literature. 
Finally, in Section~\ref{sect:conclusions} we present our conclusions. 
Throughout the paper we use $\Lambda$CDM cosmology with $\Omega_{\Lambda}=0.70$, $\Omega_m=0.30$ and $h = H_0/100 = 0.7$. 

\section{Data}
\label{sect:data}

We have observed four out of the five known $z>6$ radio-loud quasars with NOEMA. 
J1427+3312, J1429+5447 and P172+18 were observed in 2019, December 3rd, June 16th and August 14th, as part of the S19DN observing programme. 
J0309+2717 was observed in 2020, June 18th, 20th and July 20th as part of the S20CY observing programme. 
All targets were observed in band 3, which covers the range $\sim208-264$ GHz range where the redshifted [CII] emission line  ($\nu_{\rm rest}=1900.539$ GHz) falls at $z\gtrsim6$ . 
The tuning frequencies were chosen so that the [CII] emission falls in one of the side bands. 
The other side band can be used for FIR continuum emission measurements. 

The observations were carried out using configurations C, D or a combination of the two. Configuration D is the most compact and provides angular resolution $\sim1\farcs5$ ($\sim8$~kpc). Configuration C is more extended and provides resolution $\sim0\farcs8$ ($\sim4$ kpc).
Most observations were carried out with 10 antennas, except for J1429+5447 and one of the tracks for J1427+3312, for which 9 antennas were used. 
The beam sizes (0.5--1.7 arcsec) correspond to 3--10 kpc at the quasar redshifts.
The total integration time, configuration used, and synthesized beam for each quasar are listed in Table~\ref{table:observations}.

The data were calibrated using the standard calibration steps in the GILDAS software\footnote{https://www.iram.fr/IRAMFR/GILDAS}. 
We used 3C84, 3C345, 3C273 or 3C279 sources for bandpass calibration. 
The phase and amplitude were calibrated using 1417+273, 1418+546, 1147+245. 
The flux density scale was calibrated using LKHA101 for P172+18 and MWC349 for the rest of the sample. 
We flagged bad visibilities before producing the \textit{uv} tables.
We resampled all \textit{uv} tables to the resolution of 50 km~s$^{-1}$.

The dirty images were produced from the \textit{uv} tables using \textit{MAPPING} software package (part of GILDAS). 
We used natural weighting, since we expect that our targets will not be resolved and they are exactly at the center of the image where natural weighting yields the optimal sensitivity. 
The clean images were produced using the HOGBOM method \citep{Hogbom1974} in \textit{MAPPING}.

We produced continuum maps using the [CII] emission-free side band by averaging all the channels in this band. Before producing [CII] emission line maps, we averaged all the channels in the side band containing the [CII] emission to produce preliminary images. 
We found >3$\sigma$ detections in the center of the images of J1427+3312, J1429+5447 and J0309+2717. 
We then extracted spectra from the brightest pixel in this image to search for the [CII] emission line. Since no significant emission was found on the preliminary image of P172+18, we extracted spectra from the central pixel that is located at the optical position of the quasar \citep{Banados2021}.
We used the publicly available code Interferopy \citep*{interferopy} to measure the flux densities using the residual scaling correction \citep[][]{Jorsater1995_residualscaling, Walter1999_residualscaling, Novak2020}.
In this method, the flux density measured on the clean components map is corrected by the scaled residual flux density. The scaling factor is defined as the clean-to-dirty beam area ratio. This ensures consistency between the units of the residual flux density (in Jy/dirty beam) and the clean components (in Jy/clean beam).
We found >3$\sigma$ [CII] line detections for all targets where emission was detected on a preliminary image. We fit all the [CII] emission line profiles detected with a single Gaussian.
We estimated the continuum emission excluding all the channels within $\pm$1000 km~s$^{-1}$ from the peak of the [CII] line.
We subtracted the continuum using the task \textit{UV\_BASELINE} in \textit{MAPPING}.
We then produced integrated [CII] line maps using channels containing the line across 1.2$\times$FWHM (in case of a non-detection, we used FWHM=350 km~s$^{-1}$).  

Using the [CII] line maps, we determine the optimal aperture radius to extract the spectra by performing a curve-of-growth analysis. We start with an aperture size equal to the half of the semi-major axis of the beam (this is also the aperture size which we used in case of a non-detection). We chose the aperture radius at which increasing the radius changes the measured aperture integrated flux density only within $1\sigma$. In this way, we make sure that we do not miss any emission of an unresolved source  and do not introduce additional noise if we chose a bigger aperture. Similarly, we chose the aperture for measuring the continuum flux density.

\begin{table*}
\caption{NOEMA observations}
\label{table:observations}
\centering 
\begin{tabular}{c c c c c}      
\hline\hline
Quasar & $R_{4400}$\tablefootmark{a} & Integration time (hours) & Configuration & Beam size (arcsec)\\ 
\hline  
	J0309+2717 & $2500\pm500$\tablefootmark{b} & 4.8 & D & $1.7\times1.1$ \\
    J1427+3312 & $53.3\pm4.1$\tablefootmark{c} & 3.4 & CD & $1.1\times0.7$\\   
    J1429+5447 & $109.2\pm4.1$\tablefootmark{c} & 3.4 & D & $1.6\times1.4$\\
    P172+18 & $70\pm7$\tablefootmark{d} & 4.1 & C & $1.1\times0.5$ \\
    
\hline                                           
\end{tabular}
\tablefoot{
	\tablefoottext{a}{Radio-loudness $R_{4400}=S_{5GHz}/S_{4400\text{\normalfont\AA}}$ \citep{Kellermann1989}.}
	\tablefoottext{b}{\cite{Belladitta2020}}
	\tablefoottext{c}{\cite{Banados2015_rlq_fraction}}
	\tablefoottext{d}{\cite{Banados2021}}
}
\end{table*}

\section{Results}
\label{sect:results}

\begin{table*}
\caption{[CII] and underlying continuum emission measurements of the sample presented in this paper and from the radio-loud quasars at $z\gtrsim6$ in the literature}
\label{table:flux_meas}
\centering 
\begin{tabular}{c c c c c }      
\hline\hline
Quasar & $z_{[CII]}$ & $F_{cont}$  & $F_{[CII]}$  & FWHM \\ 

 &  & (mJy) & (Jy~km~s$^{-1}$) & (km~s$^{-1}$) \\ 
\hline  
J0309+2717 & $6.100\pm0.002$ & $4.39\pm0.15$  & $2.6\pm0.2$  & $240\pm20$ \\
J1427+3312 & $6.118\pm0.006$ & $0.18\pm0.05$  & $0.7\pm0.1$  & $440\pm69$ \\   
J1429+5447 & $6.190\pm0.004$  & $3.05\pm0.11$ & $3.6\pm0.2$   & $359\pm24$ \\
P172+18    & $6.823^{+0.003}_{-0.001}$\tablefootmark{a} & <0.34\tablefootmark{b} & <0.21\tablefootmark{b} 	&  -- \\
\hline      
J2318$-$3113\tablefootmark{c} & $6.4429\pm0.0003$  & $0.36\pm0.08$ & $1.52\pm0.14$   & $344\pm34$ \\
P352$-$15\tablefootmark{d} & $5.832\pm0.001$  & $0.34\pm0.04$ & $1.37\pm0.22$   & $440\pm80$ \\
\hline
\end{tabular}
\tablefoot{
	\tablefoottext{a}{This redshift is measured based on Mg\,II emission line \citep{Banados2021}.}
	\tablefoottext{b}{The limits are at 3$\sigma$ and assuming FWHM=350 km~s$^{-1}$.}
	\tablefoottext{c}{The measurements are taken from \cite{Venemans2020}}
	\tablefoottext{d}{The measurements are taken from \cite{Rojas2021}}
}
\end{table*}

\begin{figure*}
	\centering
	\includegraphics[width=1.0\textwidth]{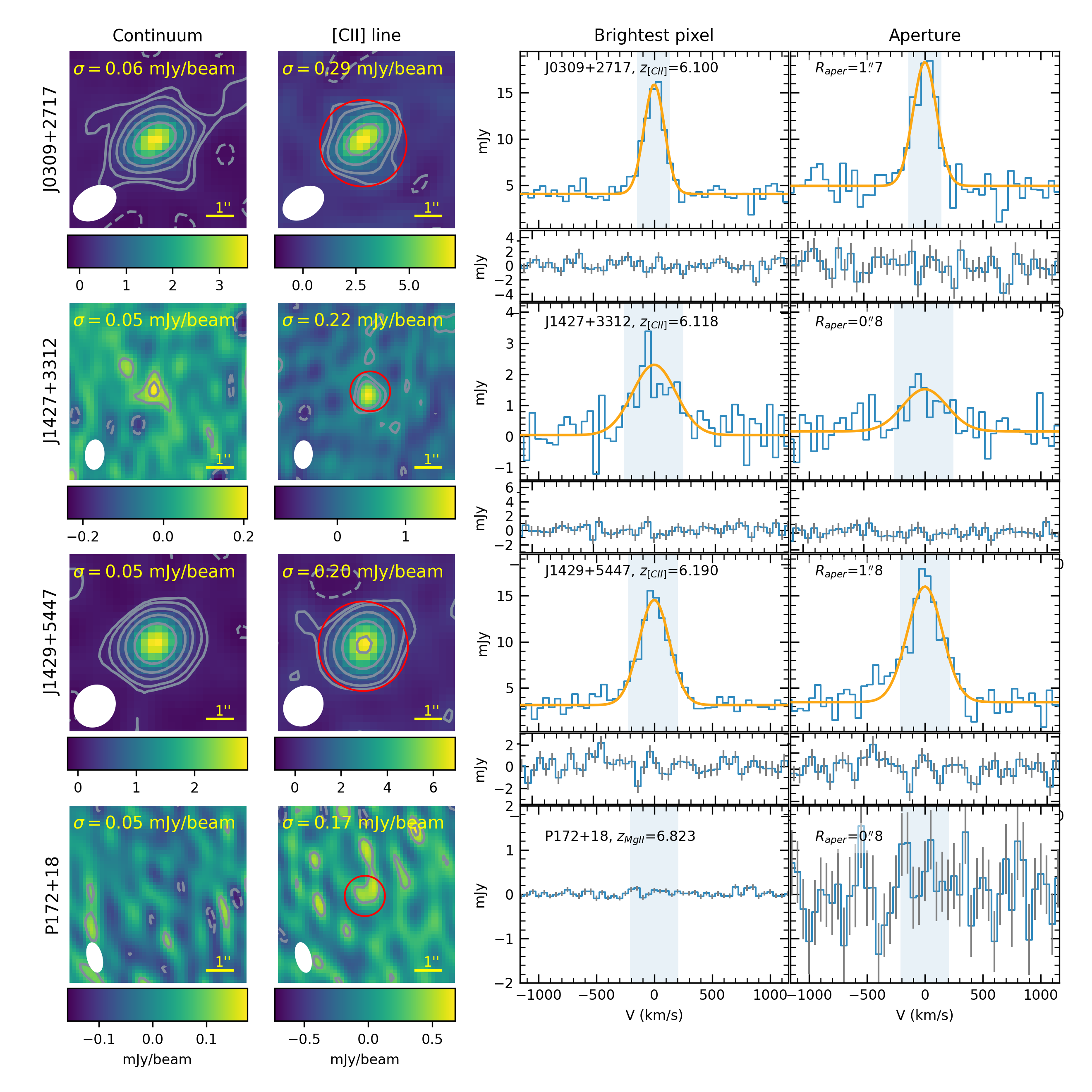}
	\caption{The images and spectra of the radio-loud quasars in our sample. 
First column: the 250 GHz continuum images. 
The contour levels are (-2, 2, 3.5, 8, 16, 32)$\times\sigma$.
Second column: The integrated [CII] line images (continuum subtracted). The contour levels are (-2, 2, 3.5, 8, 16, 32)$\times\sigma$ and $\sigma$ values are shown on the images.
The contour levels are the same as in the first column.
The red circles show the apertures, from which the spectrum in the fourth column was extracted. 
Third column: Spectra from the brightest (central) pixel. The solid orange line shows the best fit to the line with a Gaussian. The shaded area shows the channels used to create the integrated maps on the second column. Fourth column: spectra extracted using the apertures shown in images in the second column (see Section \ref{sect:data}).}
	\label{fig:summary_all}
\end{figure*}

Our analysis reveals a range of host galaxy properties of radio-loud quasars. 
The [CII] line and underlying continuum  were detected for J1427+3312, J1429+5447, J0309+2717, while they were not detected for P172+18. 
The [CII] and underlying continuum emission line measurement results are summarized in Table~\ref{table:flux_meas}. 
Fig.~\ref{fig:summary_all} shows the 250 GHz continuum and [CII] line maps and the spectra zoomed on the [CII] emission line. 
Below we discuss the results for each individual object in the sample ordered by their redshift.

\subsection{J0309+2717 ($z=6.10$)} 

J0309+2717 is the most radio-loud quasar at $z>6$. Based on its radio and X-ray properties, this source was classified as a blazar and currently it is the only blazar at $z>6$ \citep{Belladitta2020}. A one-sided jet launched by this quasar is pointing towards an observer at the angle smaller than $38\degr$ based on high resolution observations with the Karl G. Jansky Very Large Array (VLA) of the National Radio Astronomy Observatory \citep[NRAO, ][]{Spingola2020}. The jet is seen in X-ray and is extended across $\sim4\arcsec$ or $\sim20$~kpc \citep[][]{Ighina2022}. The radio spectrum of J0309+2717 is flat with a spectral index $\alpha=-0.53\pm0.02$ at frequencies from 0.147 GHz to 8.2 GHz \citep[see Fig. \ref{fig:sed_all},][]{Ighina2022,Mufakharov2020}.

J0309+2717 has the brightest 250 GHz continuum emission in our sample ($F_{\rm cont}=4.39\pm0.15$ mJy) and the second brightest [CII] line emission ($F_{\rm [CII]}=2.6\pm0.2$ Jy~km~s$^{-1}$). 
The source is unresolved.
No significant difference in shape is observed between the spectrum extracted from the brightest pixel and from an aperture with $3.4\arcsec$ diameter (see Fig. \ref{fig:summary_all}).
But the spectra extracted with an aperture has a higher flux density (by a factor of $\sim1.2$), since some of the emission is missed if a single pixel is used to extract the spectrum. Therefore, we use the aperture extracted spectrum in further analysis in Section~\ref{sect:discussion}.


\subsection{J1427+3312 ($z=6.12$)} 

J1427+3312 is the first radio-loud quasar discovered at $z>6$ \citep{McGreer2006}. After the discovery, two independent Very Long Baseline Interferometry (VLBI) follow-up studies of J1427+3312 were conducted. \cite{Momjian2008} observed a structure with two continuum components at 1.4 GHz, separated by 176 pc. \cite{Frey2008} observed J1427+3312 at 1.6 and 5 GHz and the 1.6 GHz observations revealed a double structure of this quasar with two components separated by 160 pc, comparable to the separation observed by \cite{Momjian2008}. Both these studies conclude that these two components could be the radio lobes and J1427+3312 could be a Compact Symmetric Object \citep[CSO,][]{Conway2002} with a steep radio spectrum ($\alpha=-1.1$). Based on its rest-frame ultraviolet (UV) spectrum, this source was also classified as a Broad Absorption Line (BAL) quasar \citep{McGreer2006, Shen2019}.

Our NOEMA observations reveal a [CII] line emission with $F_{[CII],int}=0.7\pm0.1$ Jy~km~s$^{-1}$. The curve of growth analysis shows that even with the small aperture the noise in the spectrum significantly increases compared to the extraction from a single pixel since the source is unresolved. We use, therefore, the spectrum extracted from the brightest pixel in the analysis in Section \ref{sect:discussion}. We use the fit with a Gaussian to this spectrum to select the channels containing the line. On the integrated image (including only the channels across 1.2$\times$FWHM), the [CII] emission is detected with a $4\sigma$ significance. In the channel corresponding to the peak of the line, the significance of the line is 8$\sigma$. The emission is unresolved at the resolution of our observations. The 250 GHz continuum has a flux density $F_{cont}=0.18\pm0.05$ mJy.


\subsection{J1429+5447 ($z=6.19$)}
\label{sect:J1429+5447}

J1429+5447 is one of the most studied in the literature objects in our sample. Radio observations of J1429+5447 cover the range from 120 MHz to 1.6 GHz \citep[see Fig. \ref{fig:sed_all},][]{Frey2011, Shimwell2019}. The radio spectrum is flat at frequencies $<5$~GHz (spectral index $\alpha\sim0.5$) but steepends at higher frequencies \citep[spectral index $\alpha=-1.0$,][]{Frey2011}. The high resolution VLBI observations of this quasar show a compact structure with a size <100 pc. It was also observed at 32 GHz \citep{Wang2011}. These observations targeted CO (2--1) emission line and they tentatively suggest the presence of a companion galaxy separated by 6.9 kpc. J1429+5447 was also observed with eROSITA and XMM-Newton and currently it is the brightest X-ray source at $z>6$ \citep{Medvedev2020, Medvedev2021}.

J1429+5447 is the second brightest quasar in our sample at $\sim$250 GHz ($F_{cont}=3.05\pm0.11$ mJy) and has the brightest [CII] emission line ($F_{[CII],int}=3.6\pm0.2$ Jy~km~s$^{-1}$). 
Interestingly, when we extract the spectrum with an aperture, thus accounting for all the flux seen in the [CII] line image,
additional flux deviant from a single Gaussian profile appears.
This may indicate that J1429+5447 is hosted in merging galaxies or that its host has AGN-driven outflows.
We will discuss this in more detail in Section \ref{sect:mergers}.

\subsection{P172+18 ($z=6.82$)}

P172+18 is the highest redshift quasar in our sample and the highest redshift radio-loud quasar known to date \citep{Banados2021}. The radio spectrum of P172+18 is steep with a spectral index $\alpha<-1.55$ \citep[see Fig. \ref{fig:sed_all},][]{Momjian2021}. It suggests that P172+18 is a Gigahertz Peaked Source (GPS). It is the only source from our sample, which is not detected in both [CII] and continuum emission (see Fig. 
\ref{fig:summary_all}). 
The $3\sigma$ upper limit for the FIR continuum flux density is $F_{cont}<0.34$ mJy.
We determine the upper limit for the [CII] flux density assuming the mean FWHM of the [CII] emission line in the host galaxies of radio-quiet quasars at $z>6$ FWHM=350~km~s$^{-1}$ \citep{Decarli2018, Venemans2020}. The $3\sigma$ upper limit with this assumption is $F_{[CII],int}<0.21$ Jy~km~s$^{-1}$.

\section{Discussion}
\label{sect:discussion}

\subsection{Are host galaxies of radio-loud quasars mergers or outflows?}
\label{sect:mergers}

\begin{table*}
\caption{The statistical criteria values for selection between the fit with one Gaussian and two Gaussians.}
\label{table:stats}
\centering 
\begin{tabular}{c c c c c}      
\hline\hline
Quasar & $\Delta\chi^2$ & p-value & $\Delta\text{\normalfont AIC}$\tablefootmark{a} & $\Delta\text{\normalfont BIC}$\tablefootmark{a} \\\hline  
J0309+2717 & 4.99 & 0.82738 & 0.47 & -10.83 \\
J1427+3312
& 0.92 & 0.17997 & -3.82 & -10.93 \\
J1429+5447 & 23.08 & 0.99996 & 10.02 & 2.91 \\
\hline      
\end{tabular}
\tablefoot{
\tablefoottext{a}{$\Delta\text{\normalfont AIC}$, $\Delta\text{\normalfont BIC}<-10$ imply that the fit with one Gaussian is preferred. $\Delta\text{\normalfont AIC}$, $\Delta\text{\normalfont BIC}>10$ imply that two Gaussians are preferred.}
}
\end{table*}

At redshifts $1<z<2.5$, 92$\%$ of radio-loud quasars are hosted in merging galaxies compared to 38$\%$ of radio-quiet quasars \citep{Chiaberge2015}. The merger fraction of radio-quiet quasars at $z\gtrsim6$ is $\sim29\%$\footnote{This value is calculated considering only the quasars which are confirmed to be radio-quiet. Including the quasars whose radio-loudness is unknown results in merger fraction 31\%. Taking into account that $\sim10\%$ of them could be radio-loud, the upper limit of the merger fraction of radio-quiet quasars is $\sim39\%$.} \citep[][]{Neeleman2021}. This suggests no evolution with redshift of the merger fraction in radio-quiet quasars. Assuming that the merger fraction in radio-loud quasars does not evolve either, most host galaxies of radio-loud quasars at $z\gtrsim6$ should be mergers. The recently identified a $z=6.44$ radio-loud quasar J2318--3113 \citep{Ighina2021} whose kinematics and morphology were studied in detail by \cite{Neeleman2021} using observations with ALMA at high spatial resolution ($\sim1$~kpc). 
They find that J2318--3113 has a disturbed morphology indicative of a recent or ongoing merger activity \citep[see Fig. 2 in][]{Neeleman2021}. The spatial resolution of our NOEMA data ($\sim6$~kpc) is not sufficient to perform such analysis. Therefore, we focus on analysing the spectral shape of the [CII] line in order to search for signs of mergers.

\begin{figure*}
	\centering
	\includegraphics[width=1.\textwidth]{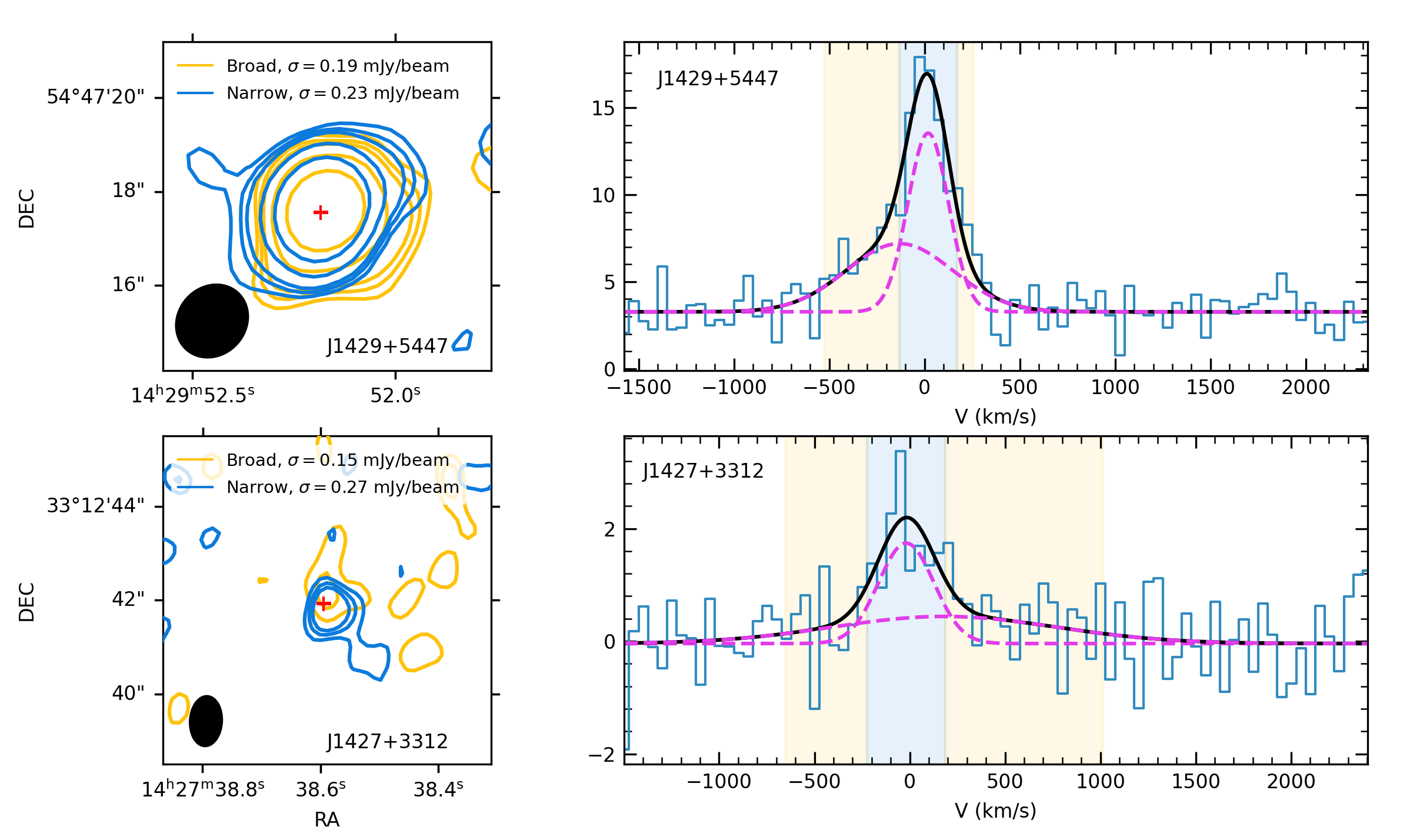}
	\caption{The images of the broad and narrow components and the spectra of J1429+5447 (\textit{top}) and J1427+3312 (\textit{bottom}).
First column: the contours on the images (blue for the narrow and yellow for the broad component). The contours correspond to (2,3,4,8,16)$\times\sigma$. 
Second column: the spectra (blue line) and the fit with two Gaussians (black solid line). The magenta dashed lines are broad and narrow components of the fit with two Gaussians. The shaded areas show the channels used to produce corresponding images of the broad and narrow components.}
	\label{fig:broadnarrow}
\end{figure*}

We noted in Section \ref{sect:results} that the [CII] spectrum of J1429+5447 shows a broader component when extracted with an aperture ($R=1\farcs77$) We fit this spectrum with two Gaussians. The two components have redshifts $z_{broad}=6.19\pm0.009$ and $z_{narrow}=6.19\pm0.003$ and FWHM$_{broad}=650\pm128$ km~s$^{-1}$ and  FWHM$_{narrow}=253\pm38$ km~s$^{-1}$. The fit with two components is preferred by both Akaike Information Criterion (AIC) and Bayesian Information Criterion (BIC, the values are listed in Table \ref{table:stats}), although the BIC difference does not exceed the threshold $\Delta\text{\normalfont BIC}>10$ for a strong significance \citep{BIC_Kass_Raftery}. The $\chi^2$ difference implies p-value of $p=0.99996$ for three additional degrees of freedom or 3.95$\sigma$ significance for a broad component.

Previously, the CO (2--1) emission of J1429+5447 was observed with a resolution of $\sim$0.7 arcsec at 32 GHz with the VLA. \cite{Wang2011} report the presence of two components of CO (2--1) emission line where the broad component is detected at 4$\sigma$ significance. The two components have redshifts $z_{east}=6.1837$ and FWHM$_{east}=400$ km~s$^{-1}$, and $z_{west}=6.1831$ and FWHM$_{west}=280$ km~s$^{-1}$. 
The widths and redshifts of the CO (2--1) line are consistent with the ones we obtained from the best fit of the [CII] line profile on the spectrum from the aperture. 
The similarity of the widths of the spectral profiles of the CO(2--1) and [CII] lines  suggests that the [CII] emission arises from the same structure as the reported CO (2--1) emission.
In that case, J1429+5447 is likely to be a merger or two gravitationally interacting sources. 

We cannot completely separate the narrow component from the broad one, but by averaging the channels within 1.2$\times$FWHM of the broad component and excluding all the channels across 1.2$\times$FWHM of the narrow component, we can obtain an image with little contamination from a narrow component. In order to obtain an image of the narrow component, we average the channels across 1.2$\times$FWHM$_{narrow}$. These channels still contain emission from the broad component, but the emission from the narrow component is comparable to the broad one or dominates it. 
We show the images for the narrow and broad components in Fig.~\ref{fig:broadnarrow}. The [CII] emitting regions of the broad and narrow component still overlap and we cannot make firm conclusions whether they could be separated as the CO(2--1) line emission observed by \cite{Wang2011}. The size of the NOEMA beam is $1.6\times1.4$ arcsec, almost two times larger than the beam of the CO(2--1) line observations and larger than the separation between the components on the CO(2--1) line map.
Therefore, higher angular resolution [CII] observations are necessary to make firm conclusions about the nature of the broad and narrow [CII] emission in this system.


We performed a similar analysis for the remaining quasars with a [CII] detection in our sample. J0309+2717 has no indication of a second component. The fit with two Gaussians results in adding a negative broad Gaussian shifted by more than 1000 km~s$^{-1}$ from the [CII] line center. This broad component does not have a physical meaning and is discarded by the BIC. The difference in AIC is not significant (see Table \ref{table:stats}). We conclude that the spectrum of J0309+2717 is represented by one Gaussian.


In Section \ref{sect:results}, we noted that J1429+3312 is unresolved and extracting the spectrum with an aperture results in a lower signal to noise ratio (SNR).
Therefore, we test the fit with one and two Gaussians on the spectrum extracted from the brightest pixel.
The fit with two Gaussians reveals a broad component of [CII] emission with FWHM$_{broad}\sim1400$ km~s$^{-1}$. The narrow component has FWHM$_{narrow}=343\pm113$ km~s$^{-1}$ consistent with the width of the fit with a single Gaussian.
The width of the broad component is consistent with what is expected for outflows.

The existence of the outflows in host galaxies of radio-quiet quasars is debated in the literature. Individual radio-quiet quasars observed at $\nu_{rest}\sim1900$ GHz predominantly have single Gaussian profiles of [CII] line with FWHM$_{[CII]}\sim350$ km~s$^{-1}$ \citep[][]{Decarli2018, Novak2020}. \cite{Maiolino2012} and \cite{Cicone2015} found such outflows in the host galaxy of J1148+5251, but later observations of this object by \cite{Meyer2022} with a larger number of antennas and the new wide-band correlator PolyFiX of NOEMA did not confirm the presence of a [CII] outflow in the host galaxy.
\cite{Decarli2018} approached the search for outflows using the stacking of ALMA [CII] observations and did not find outflow signatures. On the contrary, \cite{Bischetti2019} reports an evidence of a weak broad component with FWHM$^{broad}_{[CII]}=1730\pm210$ km~s$^{-1}$ in the stacked spectrum.
However, this is a delicate issue that might depend on different techniques employed.
\cite{Novak2020} used the stacking in the {\it uv}-plane instead of the image plane and did not find any indication for a presence of the broad component.
While the spectral profile of J1429+5447 has indication of a narrow and a broad component, the width of the broad component is much smaller than what is typical for the outflows (i.e., a few 1000s km\,s$^{-1}$).

J1427+3312 was previously classified as BAL quasar \citep{McGreer2006,Shen2019}. 
Hence, the host galaxy of this quasar can have high-velocity outflows.
We separate the broad and narrow component of J1427+3312 in the same way as described above for J1429+5447 and show the results in Fig. \ref{fig:broadnarrow}. The broad component only has a $3\sigma$ emission close to the optical position of the quasar.
The fit with two Gaussians, however, is discarded by BIC (see Table \ref{table:stats}).
The AIC neither discards nor confirms the presence of the second Gaussian.
As an additional test, we compared the fit with one and two Gaussians to the spectrum extracted with an aperture.
The broad component with the width compatible with an outflow scenario (FWHM$_{broad}\sim940$ km~s$^{-1}$) appears in the fit with two Gaussians to this spectrum as well.
Both the BIC and AIC for that fit neither discard nor confirm the broad component.
Since the [CII] emission line of J1427+3312 is rather faint and the SNR is low with the sensitivity of our observations, the presence of high-velocity outflows in J1427+3312 requires further investigation with higher sensitivity follow-up observations.

\subsection{The effects of the jet on star formation}
\label{sect:sfr}

\begin{figure*}
	\centering
	\includegraphics[]{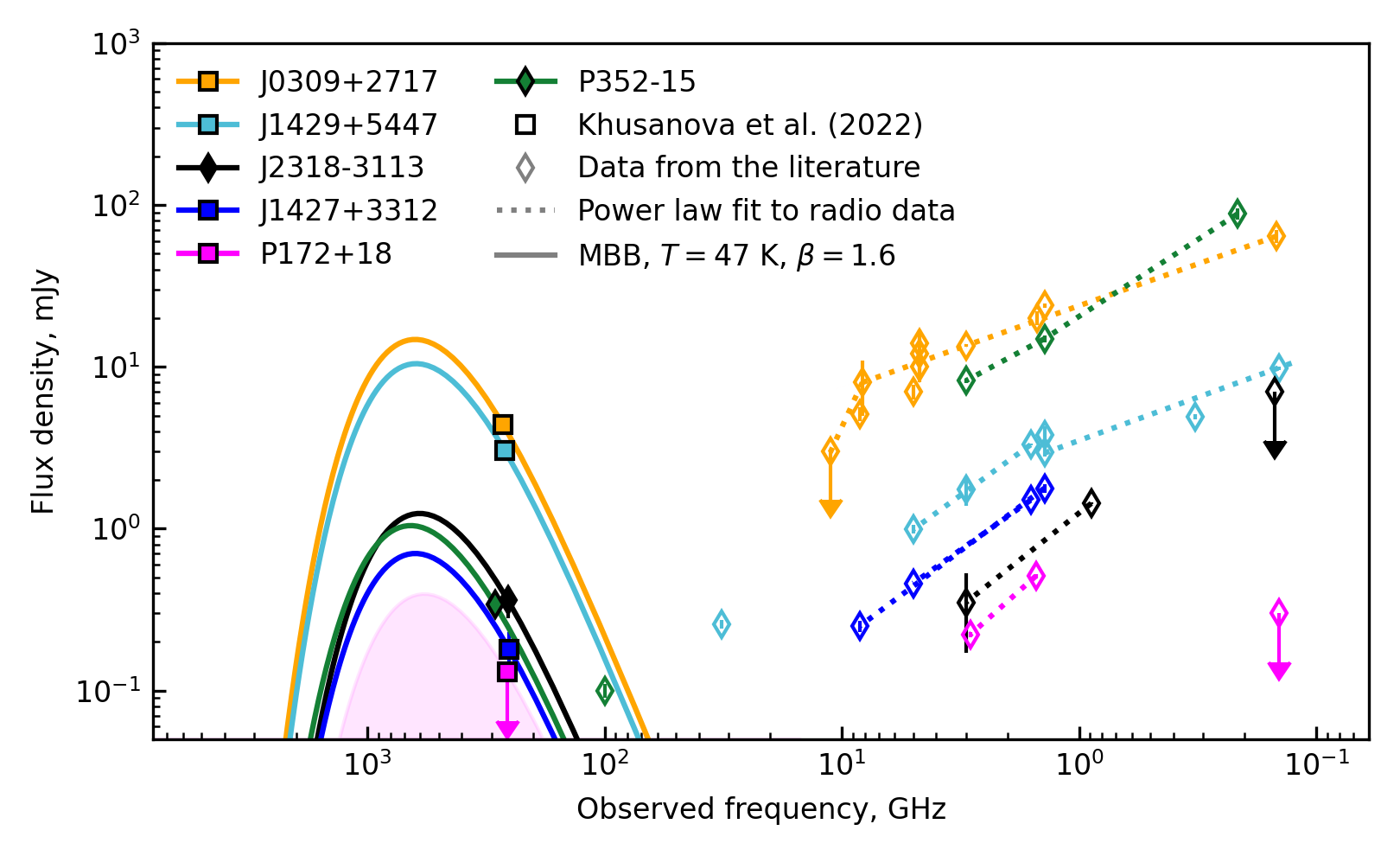}
	\caption{SEDs of all known radio-loud quasars at $z>6$ and P352--15 at $z=5.832$. 
		The filled symbols are measurements of the 250 GHz continuum flux density \citep[NOEMA measurements presented in this paper and literature data from ALMA for J2318-3113][]{Venemans2020}.	The solid lines are MBB model (with $T_{dust}=47$ and $\beta=1.6$) scaled to the observed continuum flux density. 
		The dotted lines are the power law (broken power law in the case of J0309+2717 and J1429+5447) fits to the radio data from the literature \citep{Banados2021, Condon1998, Frey2008, Frey2011,Ighina2021, Intema2017,  Momjian2008, Shimwell2019, Wang2011}. 
		The literature data are shown with thin diamonds and the NOEMA data for our sample with squares. 
		Different colors correspond to each of the radio-loud quasars as indicated in the legend. 
		The shaded area shows the region below the 3$\sigma$ limit for continuum flux density of P172+18, which is not detected with NOEMA.}
	\label{fig:sed_all}
\end{figure*}

The presence of radio-jets can enhance as well as quench star formation in the host galaxy. 
\cite{Mandal2021} suggest that both effects are present with increased SFR closer to the center of the host galaxy and decreased SFR on the outskirts. 
The IR and [CII] observations have been extensively used to estimate SFRs of host galaxies of radio-quiet quasars \citep[e.g.][]{Decarli2018, Venemans2020}. 
Their SFRs reach up to $\sim$2500~M$_{\odot}$/yr with a median of the SFR distribution at $\sim$250~M$_{\odot}$/yr. 
Here we compare these results with our measurements for the host galaxies of $z\gtrsim6$ radio-loud quasars.

To determine the IR luminosities, we use the modified black body model (MBB) and optically thin approximation \citep[][]{Beelen2006}:

\begin{equation}
	\label{eq:MBB}
	S_{\nu_{obs}}= f_{CMB} \dfrac{1+z}{D_{L}^2} \kappa_{d}(\nu_{rest},\beta) \dfrac{2h\nu_{rest}^3}{c^2} \dfrac{M_{dust}}{e^{h\nu_{rest}/k_{b}T_{dust,z}} - 1},
\end{equation}

\noindent where $f_{CMB}$ is a correction for the Cosmic Microwave Background (CMB) contrast, $D_L$ is the luminosity distance, $\kappa_{d} (\nu_{rest},\beta)=\kappa_{d(\nu_0)}(\nu_{rest}/\nu_{o})^\beta$ cm$^2$g$^{-1}$ is the opacity law, $M_{dust}$ is the dust mass, $T_{dust,z}$ is the dust temperature at given redshift, $\beta$ is the emissivity index and $\nu_{rest}=(1+z)\nu_{obs}$ is the rest frame frequency. 
For the opacity law, we assume values $\kappa_{d(\nu_0)}=2.64$ m$^2$kg$^{-1}$ at $\nu_0=c/(125 \mu m)$ from \cite{Dunne2003}. 
The dust temperature heated by the CMB is:

\begin{equation}
	\label{eq:Tdust}
	T_{dust,z} = (T_{dust}^{\beta+4} + T^{\beta + 4}_{CMB,z=0}[ (1+z)^{\beta+4}-1 ])^{ \dfrac{1}{\beta+4}},
\end{equation}

\noindent where $T_{dust}$ is the intrinsic dust temperature and $T_{CMB,z=0}$ is the CMB temperature at $z=0$ \citep{Cunha2013}. 
Here, we assume $T_{dust}=47$ K and the emissivity index $\beta = 1.6$ \citep{Beelen2006}. 
The CMB contrast is defined as:

\begin{equation}
	\label{eq:fcmb}
	f_{CMB} = 1 - \dfrac{B_{\nu_{rest}} (T_{CMB,z})}{B_{\nu_{rest}}(T_{dust,z})}
\end{equation}

\noindent where $T_{CMB,z}$ is the CMB temperature at redshift $z$ and $B_{\nu_{rest}}$ is the black body radiation spectrum. 
We scale the MBB to the continuum flux density at $\nu_{rest}\sim1900$ GHz and obtain the IR luminosity by integrating between 8 $\mu$m and 1000 $\mu$m (see Fig. 
\ref{fig:sed_all}). 
We convert the IR luminosity to SFR using the \cite{kennicutt_global_1998} relation:

\begin{equation}
	\label{eq:sfr_ir}
 SFR_{IR}=\kappa_{IR}L_{IR},
\end{equation}

 \noindent where $L_{IR}$ is the IR luminosity and $\kappa_{IR}=10^{-10}M_{\odot}$yr$^{-1}L_{\odot}^{-1}$ is a conversion factor assuming \cite{Chabrier2003} initial mass function.

We use the \cite{DeLooze2014} relation to convert [CII] luminosities to SFRs:

\begin{equation}
	\label{eq:sfr_cii}
 \dfrac{SFR_{[CII]}}{M_{\odot}yr^{-1}}	 = 3\times10^{-9} \left( \dfrac{L_{[CII]}}{L_{\odot}} \right) ^{1.18}.
\end{equation}

\noindent The [CII] luminosity is calculated as

\begin{equation}
	\label{eq:cii_lum}
	\dfrac{L_{[CII]}}{L_{\odot}}	 = 1.04\times10^{-3} \dfrac{F_{[CII]}}{\textrm{Jy km s}^{-1}} \dfrac{\nu_{obs}}{\textrm{GHz}} \left( \dfrac{D_{L}}{\textrm{Mpc}} \right) ^{2}.
\end{equation}

\noindent where $F_{[CII]}$ are the [CII] line fluxes, $\nu_{obs}$ is the observed frequency of the [CII] line and $D_{L}$ is the luminosity distance \citep[e.g.,][]{Carilli2013}.

\begin{table*}
	\caption{The IR and [CII] luminosities and SFRs of radio-loud quasars at $z\gtrsim6$ }
	\label{table:lum_and_sfr}
	\centering 
	\begin{tabular}{c c c c c}      
		\hline\hline
		Quasar &  $\log(L_{IR}/L_{\odot})$ & $\log(L_{[CII]}/L_{\odot})$ & $\log{SFR_{IR}(M_{\odot}yr^{-1})}$ & $\log{SFR_{[CII]}(M_{\odot}yr^{-1})}$ \\ 
		\hline 
J0309+2717 & $13.07\pm0.01$     & $9.40\pm0.03$     & $3.07\pm0.01$   & $2.57\pm0.04$  \\
J1427+3312 & $11.68^{+0.11}_{-0.16}$   & $8.82^{+0.06}_{-0.07}$ & $1.68^{+0.11}_{-0.15}$     & $1.88^{+0.07}_{-0.09}$ \\  
J1429+5447 & $12.92\pm0.02$     & $9.55^{+0.02}_{-0.03}$ & $2.92\pm0.02$    & $2.75\pm0.03$ \\
P172+18    & $<11.61$\tablefootmark{a} & $<8.37$\tablefootmark{a} & $<1.61$\tablefootmark{a}   & $<1.35$\tablefootmark{a} \\
J2318--3113\tablefootmark{b} & $12.02^{+0.09}_{-0.11}$   & $9.20\pm0.04$     & $2.02^{+0.09}_{-0.11}$     & $2.33\pm0.05$\\
P352--15  & $11.93\pm0.50$   & $9.09^{+0.06}_{-0.08}$   & $1.93\pm0.50$    & $2.21^{+0.08}_{-0.08}$\\
		\hline      
	\end{tabular}
	\tablefoot{
		\tablefoottext{a}{The upper limits are at 3$\sigma$}
		\tablefoottext{b}{The flux density measurements are taken from \cite{Venemans2020}. 
The luminosities and SFRs are recalculated following the approach in Section~\ref{sect:sfr}}
	}
\end{table*}

\begin{figure*}
	\centering
	\includegraphics[width=0.8\textwidth]{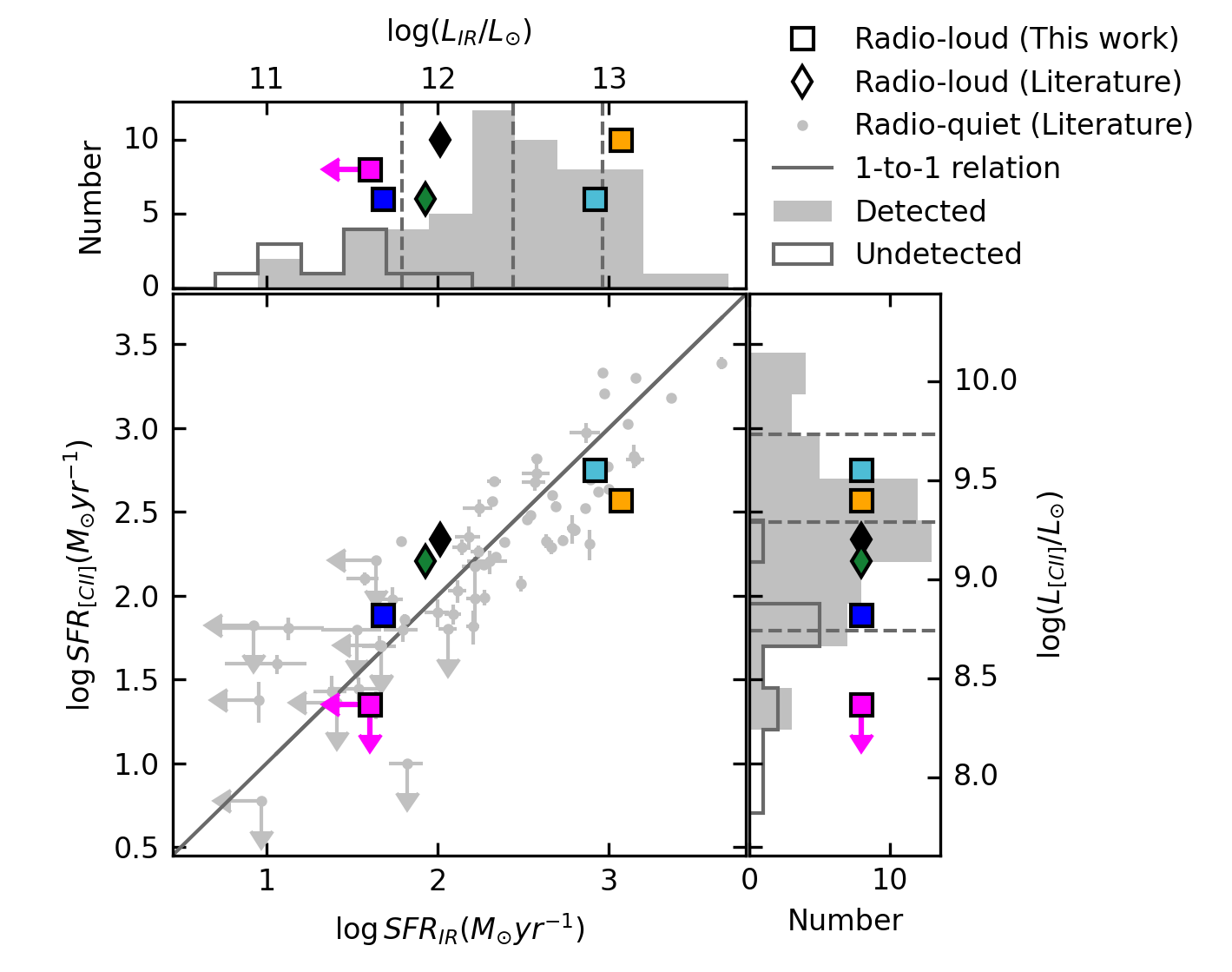}
	\caption{The SFRs of radio-quiet (grey circles) and radio-loud (colored symbols) quasars at $z\gtrsim6$ using [CII] and IR luminosities as SFR tracers. The radio-loud quasars from our sample are shown as squares and from the literature as thin diamonds. Colors for radio-loud quasars are the same as in Fig. \ref{fig:sed_all}. The solid line shows the 1-to-1 relation. The SFR (luminosity) distributions of radio-quiet quasars are shown (grey histograms). The dashed lines show the 16th, 50th and 84th percentiles of the distributions for radio-quiet quasars. References for the literature measurements are in the main text, Section \ref{sect:sfr}.}
\label{fig:sfr_cii_ir}
\end{figure*}

Following this approach, we calculated [CII] and IR luminosities of all radio-loud quasars at $z>6$ and P352--15 at $z=5.832$ \citep[]{Rojas2021}. We use the data from the literature for J2318--3113 and P352--15, which were previously observed with ALMA\citep[][]{Venemans2020,Rojas2021}.
The results are summarized in Table \ref{table:lum_and_sfr}. 
In Fig.~\ref{fig:sfr_cii_ir}, we also show the SFRs of the radio-quiet quasars, for which we used as input the observed flux densities reported in the literature \citep{Andika2020,Banados2015,Decarli2017,Decarli2018,Eilers2020,Izumi2018,Izumi2019,Maiolino2005,Venemans2012,Venemans2016,Venemans2017,Venemans2020,Walter2009,Wang2013,Wang2016,Wang2021,Willott2013,Willott2015,Willott2017,Yang2019,Yang2020}. 
We then followed a consistent approach to derive their luminosities and SFRs as described above. 

The SFR$_{IR}$ agrees well with the SFR$_{[CII]}$ with a scatter of $\sim0.3$ dex around the 1-to-1 relation for both radio-quiet and radio-loud quasars. 
The [CII] luminosities and SFR$_{[CII]}$ of the radio-loud quasars are in good agreement with the distribution for radio-quiet quasars. 
This is, however, not the case for IR luminosities and SFR$_{IR}$. 
The faintest radio-loud quasars fall all below the 21st percentile of the distribution of radio-quiet quasars, while the brightest are above $\sim$81st percentile. 
No radio-loud quasar falls within 0.65$\sigma$ from the median of the SFR$_{IR}$ distribution of radio-quiet quasars. 
This could be an indication of a bimodal distribution of IR luminosities of radio-loud quasars or simply a result of a small number statistics.

We use two sample Kolmogorov-Smirnov (KS) test to determine, whether the SFRs of radio-quiet and radio-loud quasars come from the same distribution.
For SFR$_{[CII]}$, the KS statistic is 0.22 and p-value is 0.94.
For SFR$_{IR}$, the KS statistic is 0.39 and p-value is 0.4. 
In both cases, the null hypothesis that the two samples are drawn from the same distribution cannot be rejected. Therefore, the SFR distribution of the radio-loud quasar hosts does not differ significantly from the radio-quiet population. This could mean that the presence of the jet does not have an effect on the SFR or both negative and positive feedback play a role resulting in comparable SFRs in the host galaxies of radio-loud quasars.

The role of negative and positive feedback can change depending on the evolutionary stage of the jet. Simulations show that initially the turbulence induced by the jet into the interstellar medium (ISM) causes a decrease of the SFR. Once the jet decouples from the disk, this effect weakens and the positive feedback becomes stronger \citep{Mandal2021}. The ages of the jet were previously estimated for two quasars in our sample. \cite{Momjian2008} estimated the age of J1427+3312 to be $\sim10^3$ yr based on the typical advance speed of CSOs and the distance between the radio lobes. P172+18 was classified as a Compact Steep Spectrum (CSS) radio source \citep{Momjian2021}. Assuming a typical advance speed for such objects, the age of the jet of P172+18 is $\sim1700$ yr. Both of these quasars with very young jets have faint [CII] and FIR continuum emission and low SFRs, consistent with the expectation that negative feedback plays a more important role for younger sources. However, it is necessary to obtain the estimates of the jet age for the remaining quasars in the sample to confirm this scenario.

We note that in deriving SFRs, we assumed that [CII] and IR luminosities are only related to the star formation in the host galaxies. 
This is a reasonable assumption for radio-quiet quasars at $z\gtrsim6$ \citep{Venemans2017, Novak2019,Pensabene2021, Meyer2022, Decarli2022}, but has not yet been tested for radio-loud ones.
If AGN-related sources of [CII] and IR emission are significant, we are overestimating the SFRs.

In our measurements of SFR$_{IR}$, we assumed that the FIR continuum emission is attributed to the cold dust only. 
However, this is not always the case for radio-loud quasars. 
\cite{Rojas2021} shows that the FIR continuum emission of P352--15, one of the most powerful radio-loud quasars known in the early Universe \citep{Banados2018_rlq}, cannot be reproduced by using a MBB model only. 
Therefore, the synchrotron emission from the jet can contribute to the FIR continuum emission.

The [CII] emission can arise from photon-dominated regions (PDR) associated with the star formation in the host galaxy or from the X-ray dominated region (XDR) where the gas is affected by the X-ray photons from the AGN or from shocks. 
The X-ray radiation heats the gas and can potentially cause negative AGN feedback on star formation. 
One way to determine whether XDR contributes to a significant fraction of the [CII] emission is by measuring [CII]/[CI] luminosity ratio.
XDR and PDR models show that CI is more abundant in XDR \citep[e.g.][]{Meijerink2007}. 
The observations of [CI] and [CII] emission lines in radio-quiet quasars point to the PDR origin of their [CII] emission \citep[][]{Venemans2017, Novak2019, Pensabene2021, Meyer2022, Decarli2022}, but the [CI] emission line has not yet been observed for any of the quasars in our sample.

Notably, the two quasars with the brightest [CII] emission in our sample are also the brightest X-ray sources known at $z\gtrsim6$ \citep{Medvedev2020,Medvedev2021}.
Their intrinsic X-ray radiation could be lower because X-ray luminosity can be enhanced by the inverse Compton scattering of the CMB photons by electrons in the jets, which is particularly important at high redshift \citep[e.g.,][]{Ighina2021,Connor2021, Medvedev2021}.  
Nevertheless, the observed high X-ray luminosities of these two quasars can be an indication of higher XDR contribution to the [CII] emission than in quasars with moderate or low X-ray luminosities. 

In addition, [CII] can also arise from shocks produced by interactions between jets and the gas in the host galaxy \citep[e.g.,][]{Appleton2018,Smirnova2019} or in mergers, where the shocks occur due to collision of gas rich galaxies \citep[e.g.,][]{Appleton2013, Peterson2018}.
All the quasars in our sample have evidence of jets.
Therefore it is plausible that a fraction of the [CII] emission observed originates from shocks.
In addition, J2318+3113 has been classified as a galaxy merger (Neeleman et al. 2021) and we have proposed J1429+5447 as another candidate where strong gravitational interactions might be happening (see discussion in  Section~\ref{sect:mergers}).
The combination of all these effects could explain the bright [CII] emission observed in these two sources.

 %
Our current data do not allow to determine the origin of FIR emission and [CII] line emission and it is possible that both 
 SFR$_{\rm IR}$ and SFR$_{\rm[CII]}$ are overestimated. 
If this is the case, the SFRs for radio-loud quasars would then all be below 64th percentile of the distribution for radio-quiet quasars (based on [CII]). 
The lack of highly star-forming host galaxies can be either due to stronger negative feedback from the jet or just a small number statistics.

\section{Conclusions}
\label{sect:conclusions}

We presented [CII] and underlying continuum observations of four (out of five known) radio-loud quasars at $z>6$ with NOEMA. Four radio-loud quasars are robustly detected in [CII] and their underlying continuum (three from our NOEMA survey and one from previous ALMA observations). While P172+18, the highest-redshift radio-loud quasar known to date ($z=6.8$), remained undetected (Figure \ref{fig:summary_all}). 

The spectral profiles of the [CII] line differ between all three detected host galaxies in our sample. The spectral profile of J1427+3312 is best described by one Gaussian. However, the fit with two Gaussians suggests a possible presence of the broad component with FWHM$\sim900-1400$\,km~s$^{-1}$, which can be associated with high-velocity outflows. The SNR of the current data is insufficient to make firm conclusions. J0309+2717 has a spectral profile well reproduced by a single Gaussian, similar to what is observed in radio-quiet quasars. J1429+5447 shows a clear signature of two components with widths FWHM$_{broad}=650\pm128$\,km~s$^{-1}$ and  FWHM$_{narrow}=253\pm38$ km~s$^{-1}$. Based on their similarity to CO (2--1) line observed with VLA \citep{Wang2011}, we conclude that the host galaxy of J1429+5447 is likely a merger. J2318--3113 is another  radio-loud quasar to be hosted in galaxy merger \citep{Neeleman2021}. This makes the fraction of mergers among host galaxies of radio-loud quasars $>40\%$ at $z>6$. Since our current data do not allow us to determine whether the host galaxies of the remaining three quasars are mergers, no conclusions can be made about the evolution of the merger fraction with redshift for radio-loud quasars.

The [CII] and IR luminosity distributions of radio-loud quasars are comparable with that of radio-quiet quasars (Figure \ref{fig:sfr_cii_ir}). 
If the [CII] emission and underlying continuum emission is only linked to the SFR in the host galaxy, the properties of host galaxies of radio-loud quasars are thus similar to the radio-quiet population covering the same range of SFRs. 
However, other mechanisms could be contributing to the  [CII] and IR emission present in radio-loud quasars. The FIR flux can include contribution from the synchrotron emission, while part of both the measured FIR and [CII] emission could originate from XDR or the shocks from interaction between the jet and the ISM.
If this is the case, the SFRs of the host galaxies are overestimated and can be lower than in radio-quiet quasars.
This would imply a negative feedback from the jet.
Verifying this hypothesis requires more measurements of the rest-frame FIR continuum and radio slope at high frequencies and multi-line data.

\begin{acknowledgements}  Based on observations carried out under project number S19DN and S20CY with the IRAM NOEMA Interferometer. 
IRAM is supported by INSU/CNRS (France), MPG (Germany) and IGN (Spain). S.R.R. Acknowledges financial support from the International Max Planck Research School for Astronomy and Cosmic Physics at the University of Heidelberg (IMPRS-HD). 
The National Radio Astronomy Observatory is a facility of the National Science Foundation operated under cooperative agreement by Associated Universities, Inc.
\end{acknowledgements} 

\bibliography{references}
\bibliographystyle{aa}

\end{document}